\documentclass[iop,twocolappendix,appendixfloats]{emulateapj}
\usepackage{apjfonts}
\usepackage{color}
\newcommand\beq{\begin{equation}}
\newcommand\eeq{\end{equation}}

\newcommand{\chandra}{\textit{Chandra}}

\newcommand{\boo}{Bo\"{o}tes}

\shorttitle{BCG Evolution}
\shortauthors{Lin et al.}

\begin{document}
%\begin{CJK*}{UTF8}{gbsn}

\title{The Stellar Mass Growth of Brightest Cluster Galaxies in the IRAC Shallow Cluster Survey}

\author{
Yen-Ting Lin\altaffilmark{1},
Mark Brodwin\altaffilmark{2},
Anthony H.~Gonzalez\altaffilmark{3},
Paul Bode\altaffilmark{4},
Peter R.~M.~Eisenhardt\altaffilmark{5},
S.~A.~Stanford\altaffilmark{6},\\
and Alexey Vikhlinin\altaffilmark{7}
}

\altaffiltext{1}{Institute of Astronomy and Astrophysics, Academia Sinica, Taipei, Taiwan; Kavli
Institute for the Physics and Mathematics of the Universe, Todai Institutes for Advanced Study, The University of Tokyo, Kashiwa, Chiba, Japan; ytl@asiaa.sinica.edu.tw}
\altaffiltext{2}{Department of Physics and Astronomy, University of Missouri, 5110 Rockhill Road, Kansas City, MO 64110} 
\altaffiltext{3}{Department of Astronomy, University of Florida, Gainesville, FL 32611}
\altaffiltext{4}{Department of Astrophysical Sciences, Princeton University, Princeton, NJ 08544}
\altaffiltext{5}{Jet Propulsion Laboratory, California Institute of Technology, Pasadena, CA 91109}
\altaffiltext{6}{Physics Department, University of California, Davis, CA 95616}
\altaffiltext{7}{Harvard-Smithsonian Center for Astrophysics, Cambridge, MA 02138}

%\end{CJK*}

%%%%%%%%%%%%%%%%%%%%%%%%%%%%%%%%%%%%%%%%%%%%%%
%%%%%%%%%%%%%%%%%%%%%%%%%%%%%%%%%%%%%%%%%%%%%%
\begin{abstract}

The details of the stellar mass assembly of brightest cluster galaxies (BCGs) remain an unresolved problem in galaxy formation.
We have developed a novel approach that allows us to construct a sample of clusters that form an evolutionary sequence, and have applied it to the {\it Spitzer} IRAC Shallow Cluster Survey (ISCS) to examine the evolution of BCGs in progenitors of present-day clusters with mass of $(2.5-4.5)\times 10^{14}M_\odot$.
We follow the cluster mass growth history extracted from a high resolution cosmological simulation,
and then use an empirical method that infers the cluster mass based on the ranking of cluster luminosity to select high-$z$ clusters of appropriate mass from ISCS to be progenitors of the given set of $z=0$ clusters.
We find that, between $z=1.5$ and 0.5,  
the BCGs have grown in stellar mass by a factor of $2.3$, which is well-matched by the predictions from a state-of-the-art semi-analytic model.  Below $z=0.5$ we see hints of differences in behavior between the model and observation.

\end{abstract}

\keywords{galaxies: clusters: general --- galaxies: elliptical and lenticular, cD --- galaxies: luminosity function, mass function --- galaxies: evolution}

%%%%%%%%%%%%%%%%%%%%%%%%%%%%%%%%%%%%%%%%%%%%%%
%%%%%%%%%%%%%%%%%%%%%%%%%%%%%%%%%%%%%%%%%%%%%%
\section{Introduction}
\label{sec:intro}

In a universe dominated by cold dark matter, structures are expected to grow hierarchically \citep{springel05}.  Taken at face value, such a structure formation scenario suggests that the most massive galaxies should form late.  Indeed, in the semi-analytic model (SAM) of \citet{delucia07}, the mass assembly of BCGs--the most massive galaxies in the universe--occur relatively late, in the sense that typical BCGs acquire 50\% of their final mass at $z<0.5$ through galactic mergers.
Although there is ample evidence of mergers involving BCGs at low redshifts \citep[e.g.,][]{lauer88,rines07,tran08,lin10},
the importance of stellar mass growth at late times remains unclear.
Using deep near-IR data to infer the stellar mass of BCGs across wide redshift ranges, 
it was suggested that BCGs in massive clusters have attained high stellar mass and exhibited little change in mass since $z\sim 1$ \citep{collins09,stott10}.
Using the correlation between the BCG stellar mass and cluster mass, \citet{lidman12} found that the BCGs have grown by a factor of 1.8 between $z=0.9$ and $z=0.2$, from a large sample of X-ray luminous clusters.

Some of the contradicting results may arise from inconsistent cluster sample selection (e.g., drawing different cluster samples at different redshifts that do not have any evolutionary links), and some may be due to incompatible comparisons between observations and theories 
(e.g., while theoretical models predict ``total'' magnitudes for galaxies, it is particularly difficult observationally to measure such a quantity for BCGs; \citealt{whiley08}).
Ideally, one would like to identify a cluster sample that forms an evolutionary sequence, that is, the higher-$z$ clusters are expected to be the progenitors of lower-$z$ clusters in the same sample.
With such a sample, one could then meaningfully follow the evolution of the galaxy populations, including the mass assembly of BCGs.  This approach also facilitates more direct comparisons with theoretical models.

In this paper we attempt to construct such a cluster sample and study the evolution of the stellar mass content of the BCGs,
using a subset of a complete cluster sample drawn from the {\it Spitzer} ISCS (\citealt{eisenhardt08}, hereafter E08).
We will show that the observed growth is similar to that predicted
by the Millenium Simulation \citep{springel05,guo11} at $z=0.5-1.5$, but the two disagree at $z<0.5$ at the $2\sigma$ level.

In section \ref{sec:data} we describe our cluster sample, and the numerical simulations used in this analysis.
For the construction of a cluster sample that represents an evolutionary sequence, the knowledge of cluster mass is critical.
We have developed a method to infer cluster mass from the total 4.5\micron\  luminosity of the clusters (section \ref{sec:lm}).
We then proceed to use two methods that rely on the dark matter halo merger history to infer the BCG mass growth in progenitors of present-day clusters with mass of 
$(2.5-4.5)\times 10^{14}M_\odot$, 
and compare the results to SAMs based on the Millennium Simulation (section \ref{sec:result}).
We conclude in section \ref{sec:summary}.

Throughout this paper we adopt a {\it WMAP5} \citep{komatsu09} $\Lambda$CDM 
cosmological model where $\Omega_M=1-\Omega_\Lambda=0.26$,
$H_0=71\,{\rm km\,s}^{-1} {\rm Mpc}^{-1}$, and the normalization of the matter power spectrum $\sigma_8=0.8$.

%%%%%%%%%%%%%%%%%%%%%%%%%%%%%%%%%%%%%%%%%%%%%%
%%%%%%%%%%%%%%%%%%%%%%%%%%%%%%%%%%%%%%%%%%%%%%
\section{The Data}
\label{sec:data}

%%%%%%%%%%%%%%%%%%%%%%%%%%%%%%%%%%%%%%%%%%%%%%
\subsection{Cluster Sample}
\label{sec:cls}

The cluster sample we use is from the ISCS, which consists of 335 4.5\micron-selected systems out to $z\sim 2$ over the 8.5 deg$^2$ ``Bo\"{o}tes field''.
Accurate photometric redshifts (photo-$z$) with full probability distributions $p(z)$ based on $B_WRI$[3.6][4.5] photometry are used to construct galaxy density maps in thin redshift slices \citep{brodwin06}, and clusters are detected as overdensities via a wavelet analysis.  
The resulting sample has achieved high purity based on extensive spectroscopic follow up and comparison with mock catalogs.

Despite the extensive spectroscopic campaign from the AGES survey \citep{kochanek12} and our own follow up efforts \citep[e.g., E08][]{stanford05,brodwin06,brodwin11}, 
for the majority of the high-$z$ clusters we are not able to measure their mass from the velocity dispersion.  
With the exception of two $z>1.4$ clusters described in \citet{brodwin11}, the existing X-ray data from \chandra\ is only sufficient for deriving cluster masses via X-ray scaling relations for low-$z$ clusters.  Therefore, for most of our clusters we do not have reliable mass estimates.
In this paper, we rely on the luminosity ranking method, to be described in section \ref{sec:lm}, to infer cluster mass.

The cluster photo-$z$ ($z_{\rm cl}$) is obtained from the peak of the summed $p(z)$ of candidate member galaxies within 1 Mpc, with typical accuracy of $<0.03(1+z)$ (for details see E08). % and Brodwin et al.~in prep).  
Here candidate members are defined as galaxies whose integrated photo-$z$ probability distribution within $z_{\rm cl} \pm 0.06(1+z_{\rm cl})$ is greater than 0.3.  
The total luminosity $L_{\rm tot}$ and galaxy number $N_{\rm gal}$ of the clusters are measured by subtracting the ``field'' contamination from the values obtained from the candidate members within 0.8 Mpc of the cluster center, 
where the contribution from field galaxies is measured by selecting galaxies in a manner identical to the candidate members, except in an annulus of radii $5<r<7$ Mpc around the cluster positions.

The BCGs are identified as the most luminous member galaxy in each cluster at $4.5\mu$m (i.e.~the most massive).
We have measured the BCG luminosity $L_{\rm bcg}$ from 4.5\micron\ fluxes, corrected to a 32 kpc diameter aperture.  This choice of aperture size is to ensure that we capture most of the BCG luminosity ($\sim 90\%$; \citealt{gonzalez05}).
Similar to E08, we cast both $L_{\rm tot}$ and $L_{\rm bcg}$ in unit of the passive evolving $L^*$,
based on a \citet[][hereafter BC03]{bruzual03} single burst model (formed at $z=2.5$ with the Chabrier initial mass function (IMF) and solar metallicity) that can reproduce the redshift evolution at $z\la 1.5$ of $L^*$ of cluster galaxies (see \citealt{mancone10}).

%%%%%%%%%%%%%%%%%%%%%%%%%%%%%%%%%%%%%%%%%%%%%%
\subsection{Numerical Simulations}
\label{sec:sim}

To provide guidance on the hierarchical structure formation in $\Lambda$CDM, and to estimate cosmic variance, we use two sets of large $N$-body simulations.  The first, a lightcone simulation that covers an octant of the sky up to $z\sim 3$ \citep[described in][]{sehgal10}, is capable of resolving halos with friends-of-friends mass $\ge 10^{13} M_\odot$.
The second one is of much higher resolution ($1024^3$ particles in a $320^3 h^{-3}$\,Mpc$^3$ box, hereafter referred to as the ``hi-res'' run), from which we can extract the merging and growth history of halos and subhalos, with the limiting virial mass of $6.30\times 10^{11} M_\odot$ and $3.15\times 10^{11} M_\odot$, respectively.
Both simulations were run with the {\it WMAP5} cosmology.
We will make use of the lightcone simulation in sections \ref{sec:lm} \& \ref{sec:result}, and the hi-res run in section \ref{sec:result}.

%%%%%%%%%%%%%%%%%%%%%%%%%%%%%%%%%%%%%%%%%%%%%%
%%%%%%%%%%%%%%%%%%%%%%%%%%%%%%%%%%%%%%%%%%%%%%
\section{Luminosity--Mass Relation and Halo Mass Ranking}
\label{sec:lm}

Absent traditional cluster mass proxies such as X-ray observables, weak gravitational lensing, or velocity dispersion, the simplest way to estimate a cluster's mass
is via its luminosity/stellar content or $N_{\rm gal}$ (e.g., \citealt{yang09,rozo09}).
Assuming a monotonic relationship between the mass and luminosity is unrealistic, however, because of the non-negligible scatter in the luminosity--mass correlation \citep[][hereafter L04]{lin04}.
However the {\it median} mass of the $N$ most luminous clusters (or ``top $N$'' hereafter) should have considerably lower scatter.  Here we make use of mock cluster catalogs to derive a ``lookup table'' that tells us the median mass of a sample of clusters that is rank ordered by $L_{\rm tot}$.

Our basic procedure is as follows: (1) extract a patch of the sky whose area is the same as the \boo\ field from the lightcone simulation, (2) assign a luminosity to each of the dark matter halos, (3) rank the halos by luminosity and produce the lookup table.  In practice, however, we need to take several complications into account,
such as the uncertainties of the slope $s$, scatter $\sigma_s$, and redshift evolution of the luminosity--mass correlation (hereafter $L$--$M$ relation).
Another consideration is that the mass $M_{200}$ for the halos is measured within $r_{200}$ (the radius within which the mean overdensity is 200 times the critical density of the universe at the cluster redshift),
while for our \boo\ cluster sample $L_{\rm tot}$ is measured within a metric radius of 0.8 Mpc.
We thus need a model for the spatial distribution of galaxies within clusters, and we assume the galaxies follow the \citet[][]{navarro97} profile with concentration of $c$ and scatter of $\sigma_c$ (see e.g., L04).

For $s$ and $\sigma_s$, we assume 
the possible values follow a Gaussian distribution with the mean corresponding to the observed values
based on a sample of 93 clusters at $z\sim 0$ ($s=0.85$, $\sigma_s=0.15$, e.g., L04), and the standard deviation of 0.05.  
It was found that once the luminosity is measured with respect to the evolving $L^*$, the $L$--$M$ relation does not show strong hints of redshift evolution \citep{lin06}.  We therefore assume the redshift evolution of these parameters to be negligible.\footnote{Doubling the assumed range of $\sigma_s$ at $z>1$ does not change our results.}
As for the concentration and its scatter, we note that the absolute value of $c$ does not matter for the luminosity ranking; rather, it is the ratio of $\sigma_c/c$ that perturbs the luminosity ranking.  As this ratio has not been observationally determined, we simply assume $\sigma_c/c=0.1$, 0.2, and 0.3, with equal probability.  Our results are  robust against different choices of this ratio, however.

\begin{figure}
\epsscale{1}
\plotone{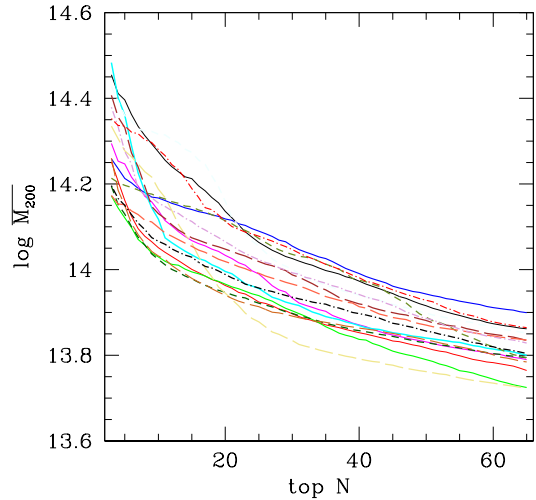}
%\vspace{-4mm}
\caption{ 
Examples of our ``lookup tables'' at $z=0.8-1.0$. 
Each curve represents the median cluster mass $M_{200}$ of the top $N$ most luminous clusters as a function of $N$ for a different mock \boo\ patch extracted from the lightcone simulation. 
}
\label{fig:cv}
\end{figure}

We have extracted 16 \boo-like patches from the lightcone. For each patch and combination of $s$, $\sigma_s$, and $\sigma_c/c$, in 
6 redshift bins ($z=0.2-0.4$, $0.4-0.6$, $0.6-0.8$, $0.8-1.0$, $1.0-1.2$, $1.2-1.5$), 
we generate 500 Monte Carlo realizations of the mock cluster catalog, and combine the results to produce a lookup table that marginalizes over the parameter uncertainties.
Fig.~\ref{fig:cv} is an illustration of the lookup tables for the 16 patches at $z=0.8-1.0$.
It is clear that cosmic variance is large: at a fixed $N$, the inferred $M_{200}$ could differ by a factor of 1.6.

%%%%%%%%%%%%%%%%%%%%%%%%%%%%%%%%%%%%%%%%%%%%%%
\section{Estimation of BCG Stellar Mass Growth}
\label{sec:result}

Our primary goal is to follow the evolution of BCGs in a sample of clusters that is believed to form an evolutionary sequence.\footnote{During the course of evolution of a cluster, the identity of its BCG may switch from one galaxy to another.  The BCGs identified by our methods are the most massive galaxies in the most massive progenitors of the $z=0$ clusters, and may not be the direct progenitors of the $z=0$ BCGs.}
We first consider a simple approach that combines the lookup tables discussed above with the knowledge of average dark matter halo growth (section \ref{sec:mergerhistory}).
We then take into account the stochastic nature of halo merger history, and show that the two methods produce very similar results (section \ref{sec:complicated}).  We further compare the results with theoretical models (section \ref{sec:mil}).

%%%%%%%%%%%%%%%%%%%%%%%%%%%%%%%%%%%%%%%%%%%%%%
\subsection{Average Dark Matter Halo Growth History}
\label{sec:mergerhistory}

\begin{figure}
%\epsscale{0.65}
\epsscale{1}
\plotone{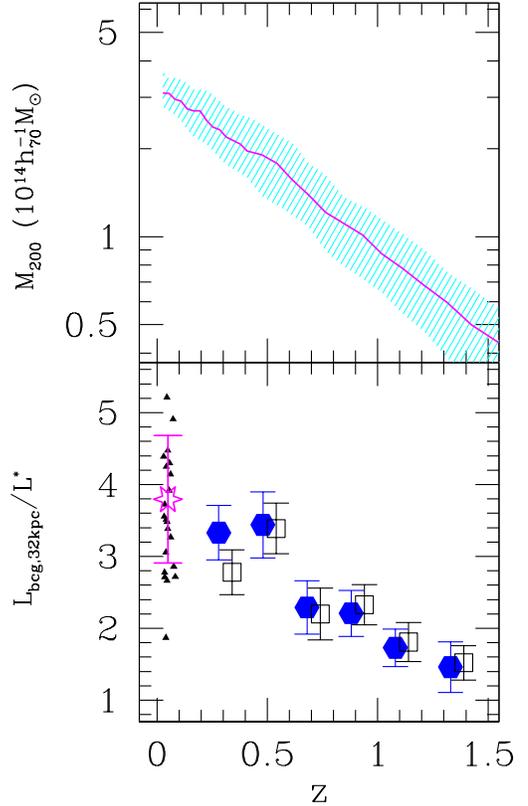}
\vspace{-8mm}
\caption{ 
  {\bf Top}: average dark matter halo mass growth (pink solid curve) and its 68\% range (shaded region), derived from the hi-res run, for halos at $z=0$ with mass of $M_{200}=(2.5-4.5)\times 10^{14}M_\odot$.
  {\bf Bottom}:  BCG luminosity evolution (solid blue points and open black squares), normalized by a passively evolving $L^*$, for ISCS clusters expected to be progenitors of present-day $M_{200}=(2.5-4.5)\times 10^{14}M_\odot$ clusters.  
The solid blue points are results based on the method described in section~\ref{sec:mergerhistory}, while the open squares are those derived from a more involved approach discussed in section~\ref{sec:complicated}.
The two methods give similar results.
  The $z\sim 0$ BCGs, taken from an enlarged version of the cluster sample presented in \citet{vikhlinin09}, are shown as small black triangles; the large magenta star represents their mean value.  The BCGs have grown by a factor of 2.7 since $z=1.5$.
}
\label{fig:simp}
\end{figure}

Using the hi-res run, we can extract the full merger history for the dark matter halos.  In Fig.~\ref{fig:simp} (top panel) we show the mass growth history of halos whose present-day mass is 
$M_{200}=(2.5-4.5)\times 10^{14}M_\odot$
(hereafter the target mass range).
The mass range is chosen  to have enough $z\sim 0$ BCGs (see below), while keeping the spread in mass of high-$z$ progenitors small.
The solid curve and the shaded region represent the median and the 68\% range spanned by the most massive progenitors, respectively.
Therefore, after specifying the mass of the $z\sim 0$ clusters, using such curves we could know the typical mass of their progenitors at any redshift.  We can then use the lookup tables of section \ref{sec:lm} to select the top $N$ most luminous clusters at the target redshift whose median mass matches the expected progenitor mass, and study the properties of BCGs in these clusters.

In Fig.~\ref{fig:simp} (bottom panel) we show as solid blue points the mean BCG luminosity $L_{\rm bcg}$ (scaled by the evolving $L^*$) within 32 kpc diameter for progenitors of present-day clusters whose mass is 
in the target mass range.
As we have scaled out the stellar aging by normalizing the luminosity to $L^*$, we could attribute the change of $L_{\rm bcg}/L^*$ as due to merger, accretion, and star formation, and will regard this quantity as a measure of stellar mass growth of BCGs.
The errorbars include the cosmic variance (estimated by the scatter in $L_{\rm bcg}$ resulted from using lookup tables from the 16 mock \boo\ patches) and a conservative 20\% systematic uncertainty to account for the 
impact of photometric redshift selection.

At $z<0.1$, the volume probed by ISCS is not large enough to contain any cluster in the
target mass range.
We thus use an enlarged version of the low-$z$ cluster sample presented in \citet{vikhlinin09}.
These 76 clusters are selected by the same criteria as described in \citet{vikhlinin09}, with the exception of a lower X-ray flux limit ($7.5\times 10^{-12}$ erg/s/cm$^2$).  All of them have high quality {\it Chandra} observations, allowing us to estimate their mass accurately via the $Y_X$--$M_{500}$ scaling relation, where $Y_X$ is the product of X-ray temperature and the intracluster medium mass \citep{kravtsov06}, and $M_{500}$ is defined analogously as $M_{200}$.  
After converting $M_{500}$ to $M_{200}$ assuming a \citet{navarro97} profile with $c=5$, there are 22 clusters within the target mass range at $z<0.1$.

The BCG luminosity for these nearby clusters within the 32 kpc diameter aperture is measured from the  Wide-field Infrared Survey Explorer (WISE, \citealt{wright10}) all-sky data release, with a redshift-dependent correction factor applied, as described in the Appendix.
We use the 3.4\micron\ data from WISE, as it is closer to the rest frame wavelength probed by the ISCS 4.5$\mu$m data.
The small black points in the Figure represent the individual BCGs, while the magenta star symbol is their mean luminosity (scaled by $L^*$).
Taking these results together, the stellar mass content of BCGs has increased by a factor of 2.7 or so since $z\approx 1.5$.

%%%%%%%%%%%%%%%%%%%%%%%%%%%%%%%%%%%%%%%%%%%%%%
\subsection{Taking Detailed Halo Merger History into Account}
\label{sec:complicated}

\begin{figure}
\epsscale{1}
\plotone{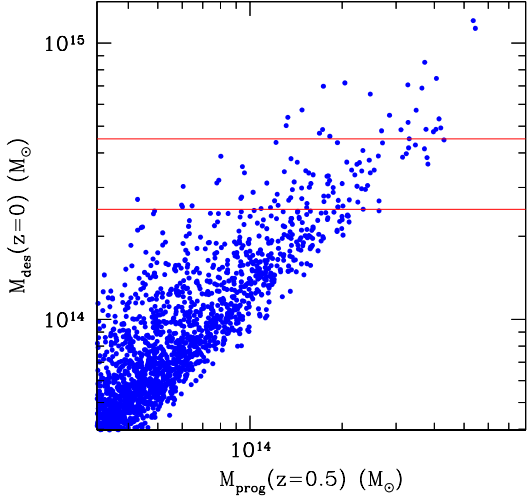}
\vspace{-3mm}
\caption{ 
  Mass of descendant and progenitor halos (at $z=0.5$) from the hi-res run.  The two horizontal lines delineate the present-day cluster mass on which we focus in this paper [$M_{200}=(2.5-4.5) \times 10^{14}M_\odot$].
}
\label{fig:hist}
\end{figure}

The approach employed in section \ref{sec:mergerhistory} ignores variations in the merger history of clusters.
In Fig.~\ref{fig:hist} we show a comparison of descendant and progenitor halo masses at $z=0$ and 0.5 from the hi-res run.
The horizontal lines delineate the target mass range of the present-day clusters used in section \ref{sec:mergerhistory}.  It can be seen that the halos at $z=0.5$ that grow into 
the target mass range
at $z=0$ span a wide range in mass 
[i.e., $(0.4-4)\times 10^{14} M_\odot$].
Here we try to take this varied degree of growth into account.

For every redshift bin considered in section \ref{sec:lm}, we need to determine two normalized probability distributions: (1) the distribution in progenitor mass $p_1(M_p|M_d)$ of those halos that will grow into the target mass range $M_d$ at $z=0$, and (2) the likelihood of a progenitor having the probable  mass [i.e., $p_1(M_p|M_d)>0$] that actually becomes a halo within the target mass range at $z=0$, $p_2(M_p,M_d)$.
In other words, for $p_1(M_p|M_d)$ we  measure the mass distribution of those halos that lie in the horizontal band in Fig.~\ref{fig:hist}, and for $p_2(M_p,M_d)$ we would like to know, for all progenitors that have the same masses as those lying in the horizontal band in Fig.~\ref{fig:hist}, the likelihood that their descendant will end up in the narrow mass range bracketed by the horizontal lines.

We measure $p_1(M_p|M_d)$ and $p_2(M_p,M_d)$ at the six redshift bins using the hi-res run.  The mean BCG luminosity is determined as
\begin{equation}
\overline{L_{\rm bcg}} = \int \int L_{\rm bcg}(M_p) p_1(M_p|M_d) p_2(M_p,M_d) dM_p dM_d,
\end{equation}
where a cluster's mass is again inferred by the lookup table and the luminosity ranking.
The BCG mass growth derived from this method is shown as open black squares in Fig.~\ref{fig:simp} (bottom panel), which are plotted alongside those points deduced by the method presented in section \ref{sec:mergerhistory} (solid blue points).  
It is reassuring to see that the two methods give  similar results.  For simplicity, hereafter we  only present the results based on the ``simpler'' approach of section \ref{sec:mergerhistory}.

%%%%%%%%%%%%%%%%%%%%%%%%%%%%%%%%%%%%%%%%%%%%%%
\subsection{Comparison with Millennium Simulation}
\label{sec:mil}

\begin{figure}
\epsscale{1}
%\plotone{fig_cftwo2.eps}
\plotone{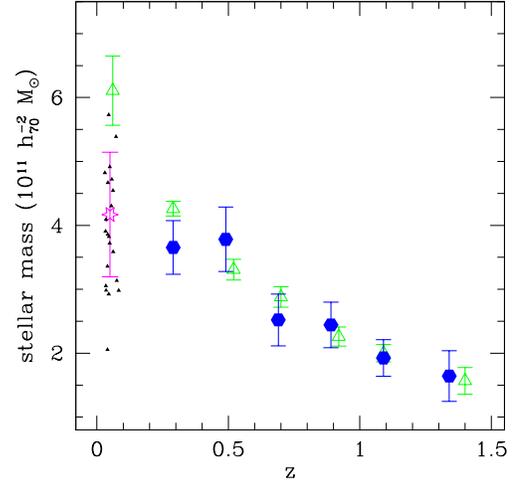}
\vspace{-3mm}
\caption{ 
Stellar mass growth of BCGs.  The solid blue points are our measurements, using the method presented in section \ref{sec:mergerhistory}.
The errorbars do not include systematic errors associated with conversion from luminosity to stellar mass.
The results from the SAM of \citet{guo11} are shown as open triangles.  The small black points show $z\sim 0$ BCGs; their mean value is shown as the pink star symbol in both panels.
The model and measurement agree well with each other at $z=0.5-1.5$; at lower redshift, the growth of model BCGs appears to increase, while 
very little growth is found for the observed BCGs.
}
\label{fig:cf}
\end{figure}

Our results show that, since $z\sim 1.5$, the BCGs in progenitors of present-day intermediate mass clusters 
have grown by a factor of $\sim 2.7$.  Here we compare these measurements with the predictions from the Millennium Simulation, based on the SAM of \citet{guo11}.

We have queried the Millennium Simulation database and found all the most massive progenitors of $z=0$ halos 
in the target mass range.
In each of the six redshift bins, 
we identify the central galaxies and measure the mean of their stellar mass.\footnote{Although the \citet{guo11} model considers contributions from the intracluster stars (ICS), we use the stellar mass that excludes the ICS component, as the sensitivities of ISCS and WISE are insufficient to detect the ICS.}
Using the most massive galaxies instead of the central ones does not change our results.
At $z\sim 0$, the aperture that encloses half of the mass in model BCGs is $\approx 27$ kpc, comparable to our chosen aperture of 32 kpc.

In practice, to take account of cosmic variance and thus make a fairer comparison with observations, for each redshift bin, we divide the Millennium Simulation box into $8-27$ sub-volumes comparable to the \boo\ observations (or the local observations, for the $z\sim 0$ bin), and adopt the weighted mean and error from all the sub-volumes for the model BCG stellar mass estimates.

The model predictions for the stellar mass growth are shown as green open triangles in Fig.~\ref{fig:cf}.
For the stellar mass of observed BCGs, we simply multiply the (time-dependent) stellar mass of the passively evolving BC03 model with the observed $L_{\rm bcg}/L^*$ ratio. 
We regard the uncertainties associated with the conversion from luminosity to stellar mass (typically $\sim 0.2$ dex) as a systematic error, which is not included in the errorbars in Fig.~\ref{fig:cf}.
Both our BC03 model and the SAM of \citet{guo11} employ the Chabrier IMF, thereby reducing one potential concern of such a comparison.
The model agrees with our measurements remarkably well, between $z=0.5$ and 1.5, where a factor of 2.3 growth is found.
However, the two disagree at the $2\sigma$ level at $z<0.5$.
While the observed BCGs show only a small increase in stellar mass content down to $z\sim 0$, the model BCGs appear to exhibit an accelerated growth below $z=0.5$; 
that is, 46\% of the final mass of the model BCGs is acquired between $z=0.5$ and 0.  The corresponding fraction for the observed BCGs is $<10\%$.

We find quantitatively very similar results when we compare to the model of \citet{delucia07}, which is also based on the Millennium Simulation.

One potential concern of our comparison with the model stems from uncertainties in the  mass estimates for the ISCS clusters.  Although our luminosity ranking method described in section~\ref{sec:lm} should allow us to reliably select clusters of the desired mass, a definitive calibration of this method awaits a formal weak lensing analysis (Lin et al.~2013, in prep.).  Here we evaluate the effect of  uncertainties in the cluster mass by first selecting halos whose present-day mass lies in the range $(1-9)\times 10^{14} M_\odot$, and then perturbing their masses with a log-normal random variable with standard deviation of $\sigma=0.3$ (except for $z\sim 0$ halos, for which $\sigma=0.08$ is chosen to reflect the much better accuracy of \chandra-based masses).  We then study the BCG growth in halos whose perturbed mass lies in the appropriate halo mass range.  The net effect is to lower the stellar mass of model BCGs, and slightly increase the discrepancy 
between the observations and model, but does not qualitatively alter our conclusions.
We conclude that the different behavior of BCG mass assembly history between the model and observation at $z<0.5$ as found above is robust against cluster mass uncertainties.

%%%%%%%%%%%%%%%%%%%%%%%%%%%%%%%%%%%%%%%%%%%%%%
%%%%%%%%%%%%%%%%%%%%%%%%%%%%%%%%%%%%%%%%%%%%%%
\section{Discussion and Conclusion}
\label{sec:summary}

Given its area and depth, the ISCS provides a unique dataset to study cluster galaxy population evolution from $z\sim 2$ to present-day.
We have developed an empirical approach that takes into account the effect of cosmic variance and overcomes the difficulty in inferring cluster masses from optical/IR cluster surveys, and have applied it
 to study the evolution of BCGs in progenitors of present-day clusters with mass of 
$(2.5-4.5)\times 10^{14}M_\odot$.
Our two methods to construct cluster samples that form an evolutionary sequence rely heavily on our knowledge of the merger history of dark matter halos (provided by numerical simulations), and both give consistent results (sections \ref{sec:mergerhistory} \& \ref{sec:complicated}).

Using a large but heterogeneous cluster sample,
\citet{lidman12} have detected a factor of 1.8 growth in BCG stellar mass between $z=0.9$ and $0.2$, which is similar to our finding.  
They have taken into account of the expected cluster mass growth between different cosmic epochs (a bit similar to our approach in section \ref{sec:mergerhistory}), and have focused on more massive clusters.
It is encouraging to see consistent results emerging from two independent analyses.
It is worth emphasizing that our method in section \ref{sec:complicated} allows us to take the stochastic merger history into account, while follow the evolution of clusters.
Our way of inferring the cluster mass via the luminosity ranking also enables us to probe a wide range in redshift, 
pushing the upper limit beyond the current capability of the X-ray surveys.

A comparison of our results with the SAM of \citet{guo11} shows good agreement between $z=1.5$ and 0.5.
At lower redshifts, there are suggestions of different behavior between the model and observation, however.
While the growth of model BCGs is accelerating at late times, that of the observed BCGs is slowing down.
Such a contrast suggests the period of $z=0-0.5$ is potentially key in differentiating models of BCG assembly history.

Our method is designed to trace the evolution of galaxies in clusters.  For the field galaxies,
\citet{vandokkum10} have studied the stellar mass growth by selecting  
galaxies at a fixed number density.  As advocated by these authors, such a selection provides a meaningful way to pick up galaxies that form an evolutionary sequence (although strictly not in the sense of dark matter halo growth), and is therefore complementary to our approach here.

Although in principle our method can be applied to study the BCG evolution in progenitors of present-day clusters of {\it any} mass range, 
extending much beyond the limited mass range presented here is beyond the capability of the ISCS cluster sample.
Studying higher mass clusters requires progenitors too massive to be found in sufficient numbers in the \boo\ field at intermediate-$z$.  The depth of our photometry and the cluster sample size at high-$z$ also prevents us from tracing lower mass present-day clusters.
With the depth and large area coverage of the upcoming Subaru HyperSuprime Cam Survey \citep{takada10} and SPT-{\it Spitzer} Deep Field (Ashby et al.~2013, in prep.), we can apply this method and study the evolution of BCGs in much greater detail,
especially for the  $z=0-0.5$ period.

%%%%%%%%%%%%%%%%%%%%%%%%%%%%%%%%%%%%%%%%%%%%%%
%%%%%%%%%%%%%%%%%%%%%%%%%%%%%%%%%%%%%%%%%%%%%%
\acknowledgments

We thank Laurie Shaw and Antonio Vale for constructing the merger trees used in this work. 
We are grateful to the anonymous referee for a report that improves the paper.
YTL thanks Gabriella De Lucia, David Spergel, and Jerry Ostriker for helpful discussions.
YTL acknowledges supports from the National Science Council grant NSC 102-2112-M-001-001-MY3, as well as
WPI Research Center Initiative, MEXT, Japan, during the course of this work.
This work was supported by National Science Foundation grants
AST-0707731 and AST-0908292.
Computer simulations and analysis were supported by the
NSF through resources provided by XSEDE
and the Pittsburgh Supercomputing Center,
under grant AST070015; computations were also
performed at the TIGRESS high performance
computer center at Princeton University, which is jointly supported by
the Princeton Institute for Computational Science and Engineering and
the Princeton University Office of Information Technology.
This work is based in part on observations made with the Spitzer Space Telescope, which is operated by the JPL/Caltech under a contract with NASA.
This publication makes use of data products from WISE, a joint project of UCLA and JPL/Caltech, funded by NASA.
The Millennium Simulation databases 
were constructed as part of the activities of the GAVO.

%%%%%%%%%%%%%%%%%%%%%%%%%%%%%%%%%%%%%%%%%%%%%%
%%%%%%%%%%%%%%%%%%%%%%%%%%%%%%%%%%%%%%%%%%%%%%

\appendix

\section{Difference in BCG photometry between WISE and IRAC}

\begin{figure}
%\epsscale{0.7}
\plotone{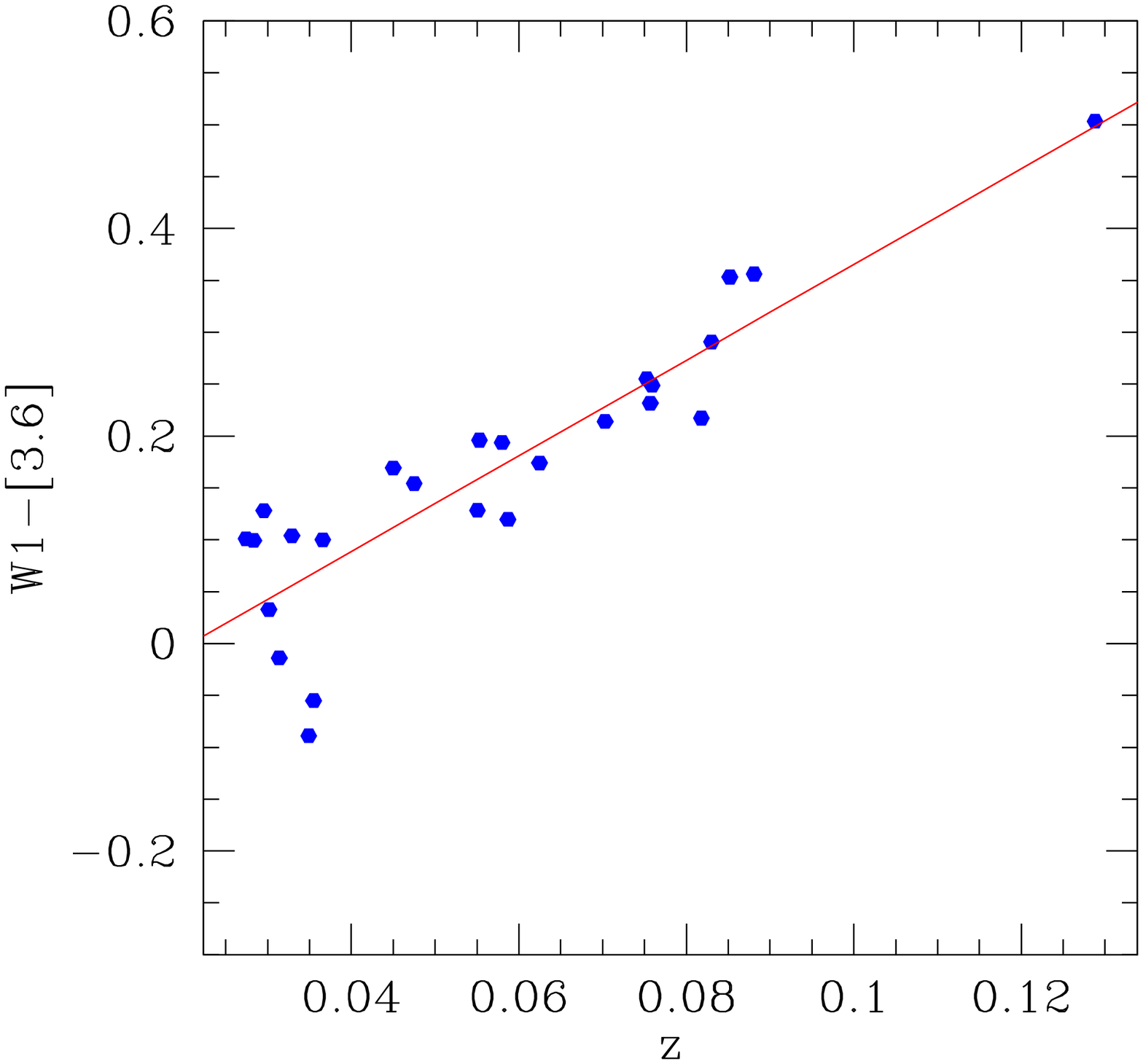}
%\vspace{-15mm}
\caption{ 
 Difference in magnitude between WISE channel 1 (W1) and IRAC channel 1 ([3.6]), measured within 32 kpc diameter with SExtractor, for a sample of 25 BCGs available in the {\it Spitzer} Heritage Archive.  The solid line shows the least squares fit to the data (see Equation A1).
}
\label{fig:wise}
\end{figure}

As the resolution of WISE channel 1 (W1) is low (6.1\arcsec) compared to IRAC, and the filter response function is somewhat different from IRAC channel 1 ([3.6]), we have compared the aperture photometry between W1 and [3.6] for a sample of 25 nearby BCGs and derived the difference in flux within the 32 kpc aperture.
More specifically, we have searched the {\it Spitzer} Heritage Archive for well-known X-ray clusters whose BCG can be unambiguously identified, and the query returned 25 clusters.  We have then measured the photometry within several apertures (up to 60\arcsec diameter) on both the IRAC images and WISE atlas images using SExtractor \citep{bertin96},
and interpolated the measurements to infer the magnitude within 32 kpc diameter (in Vega system). Fig.~\ref{fig:wise} shows the difference in magnitude (W1$-$[3.6]) for the BCGs as a function of redshift.  The trend is mainly driven by the difference in resolution between the two instruments.
A least squares fit to the data points yields
\begin{equation}
{\rm W1}-[3.6] = 4.5z-0.09.
\end{equation}
We have applied the correction factor thus inferred to the WISE photometry for the BCGs in the \citet{vikhlinin09} sample.

%%%%%%%%%%%%%%%%%%%%%%%%%%%%%%%%%%%%%%%%%%%%%%
%%%%%%%%%%%%%%%%%%%%%%%%%%%%%%%%%%%%%%%%%%%%%%

\end{document}